\documentclass[reprint,amsmath,amssymb,aps,pra,showpacs]{revtex4-1}
\usepackage{caption3}
\usepackage{keyval}
\usepackage{subcaption}
\usepackage[hidelinks,linkcolor = blue,citecolor = blue]{hyperref}
\usepackage{color}
\usepackage{graphicx}
\usepackage{dcolumn}

\begin{document}


\title{Cavity enhanced second order nonlinear photonic logic circuits}
\author{Rahul Trivedi and Uday K. Khankhoje}
\affiliation{Electrical Engineering, Indian Institute of Technology Delhi, Hauz Khas, New Delhi, India -- $110016$}
\author{Arka Majumdar}
\email{arka@u.washington.edu}
\affiliation{Electrical Engineering and Physics, University of Washington, Seattle, WA -- $98195$}
\begin{abstract}
A large obstacle for realizing quantum photonic logic is the weak optical nonlinearity of available materials, which results in large power consumption. In this paper, we present the theoretical design of all-optical logic with second order ($\chi^{(2)}$) nonlinear bimodal cavities and their networks. Using semiclassical models derived from the Wigner quasi-probability distribution function, we analyze the power consumption and signal-to-noise ratio (SNR) of networks implementing an optical AND gate and an optical latch. Comparison between the second and third order $(\chi^{(3)})$ optical logic reveals that while the $\chi^{(3)}$ design outperforms the $\chi^{(2)}$ design in terms of the SNR for the same input power, employing the $\chi^{(3)}$ nonlinearity necessitates the use of cavities with ultra high quality factors ($Q\sim 10^6$) to achieve gate power consumption comparable to that of the $\chi^{(2)}$ design at significantly smaller quality factors ($Q \sim 10^4$). Using realistic estimates of the $\chi^{(2)}$ and $\chi^{(3)}$ nonlinear susceptibilities of available materials, we show that at achievable quality factors ($Q \sim 10^4$), the $\chi^{(2)}$ design is an order of magnitude more energy efficient than the corresponding $\chi^{(3)}$ design.
\end{abstract}
\pacs{42.79.Ta, 42.50.Pq, 42.50.Lc}
\maketitle

\section{\label{sec:intro}Introduction}

All optical signal processing can potentially reduce propagation delays across the network due to increased speeds of signal propagation ($\sim 10^8$ m/s) as compared to electronic signal processing platforms which are limited by the saturation velocity of electrons in the circuit ($\sim 10^5$ m/s). Systems performing signal processing directly on optical signals can thus achieve significantly higher speeds at a lower power compared to their microelectronic counterparts \cite{cotter1999nonlinear,miller2010optical}.

The primary difficulty in implementing an all-optical signal processor lies in achieving optical nonlinearity at low energy levels. Conventional bulk optical devices operate at very high input power levels due to weak optical nonlinearities of most materials. Sustained confinement of optical energy to small volumes using optical cavities can reduce the optical power consumption due to increased light-matter interaction, thereby making the devices more energy efficient. In fact, with strong spatial and temporal confinement of light, one can reach truly quantum optical regime \cite{imamoglu1997strongly,majumdar2013single}. Recent advances in micro- and nano-fabrication \cite{chen2015nanofabrication,utke2012nanofabrication} have enabled the large-scale on-chip integration of resonators and other linear optical elements (directional couplers, phase shifters). Hence, this is an opportune time to revisit the problem of optical logic using nonlinear quantum optical devices.

Recently, researchers have proposed digital optical systems that work at intermediate photon numbers ($\sim 100$ photons) with networks of $\chi^{(3)}$ cavities \cite{santori2014quantum,mabuchi2011nonlinear}. Even though it is possible to fabricate high quality $\chi^{(3)}$ nonlinear silicon cavities \cite{dinu2003third,bristow2007two,niehusmann2004ultrahigh,deotare2009high}, significantly stronger nonlinearity can be achieved by employing $\chi^{(2)}$ nonlinear III-V systems such as Gallium Arsenide (GaAs) or Gallium Phosphide (GaP) \cite{shoji1997absolute,bergfeld2003second}, with experimentally achievable quality factors of $\sim 10^4$  \cite{sweet2010gaas,combrie2008gaas}. Using this second order nonlinearity, one can potentially reduce the overall energy consumption of optical circuits. We note that the thermo-optic effect \cite{almeida2004optical}, carrier injection \cite{nozaki2012ultralow,nozaki2013ultralow,nozaki2010sub} or optoelectronic feedback \cite{sodagar2015optical,majumdar2014cavity} can also be used to implement low-power optical logic. However, such systems rely on carrier generation and hence their response time is ultimately limited by the carrier diffusion rate. Besides, the generated carriers add significant amount of noise to the system output \cite{bonani1999generation,vahala1983semiclassical}, making the overall signal-to-noise ratio (SNR) low at low optical power.  Employing optical nonlinearities, such as $\chi^{(2)}$  or $\chi^{(3)}$ nonlinearity can provide significantly larger speeds of operation in addition to avoiding the noise originating from the generated carriers. 
 
In this paper, we investigate the design of optical logic with $\chi^{(2)}$ cavities. By comparing the performance of the $\chi^{(2)}$ and $\chi^{(3)}$ designs in terms of gate power consumption, we show that the $\chi^{(2)}$ design is a more energy efficient alternative to the $\chi^{(3)}$ design at achievable quality factors ($Q\sim 10^4$). An analysis of gate SNR shows that while the $\chi^{(3)}$ design has a better noise-performance as compared to the $\chi^{(2)}$ design for the same input power, operating the $\chi^{(3)}$ design at an input power comparable to that required by the $\chi^{(2)}$ design (with $Q \sim 10^4$) requires ultra high cavity quality factors ($Q \sim 10^6$). 

An accurate analysis of photonic circuits needs to incorporate the effect of quantum noise on their performance. Several exact formalisms, such as the SLH formalism (S,L and H refering to scattering, collapse and hamiltonian operators respectively) \cite{gough2009series}, have been developed to analyze a large quantum network. However, a full quantum optical simulation of a network consisting of large number of cavities is computationally intractable. For an approximate but computationally efficient analysis of the optical networks, we use a semiclassical model based on the Wigner quasi-probability distribution function \cite{carter1995quantum,mandel1995optical} wherein the operators are modeled as stochastic processes and the quantum nature of the optical fields is approximated by an additive noise over a mean coherent field. The analysis that we use in this paper is similar to the analysis in \cite{santori2014quantum} with $\chi^{(3)}$ photonic circuits.

{{This paper is organized as follows: In section \ref{sec:single_cavity} we use the Wigner quasi-probability to analyze the characteristics of a single nonlinear cavity. The threshold power achievable with $\chi^{(2)}$ nonlinearity is compared to that achieved with $\chi^{(3)}$ nonlinearity for different cavity quality factors. In section \ref{sec:Optical logic networks}, we present the design of an optical AND gate and an optical latch with $\chi^{(2)}$ nonlinearity and compare its noise performance and power consumption with that achieved by employing $\chi^{(3)}$ nonlinearity.}}

\section{Nonlinear cavity characteristics\label{sec:single_cavity}}
\begin{figure}[b]
\centering
\begin{subfigure}{0.32\textwidth}
\includegraphics[scale=0.31]{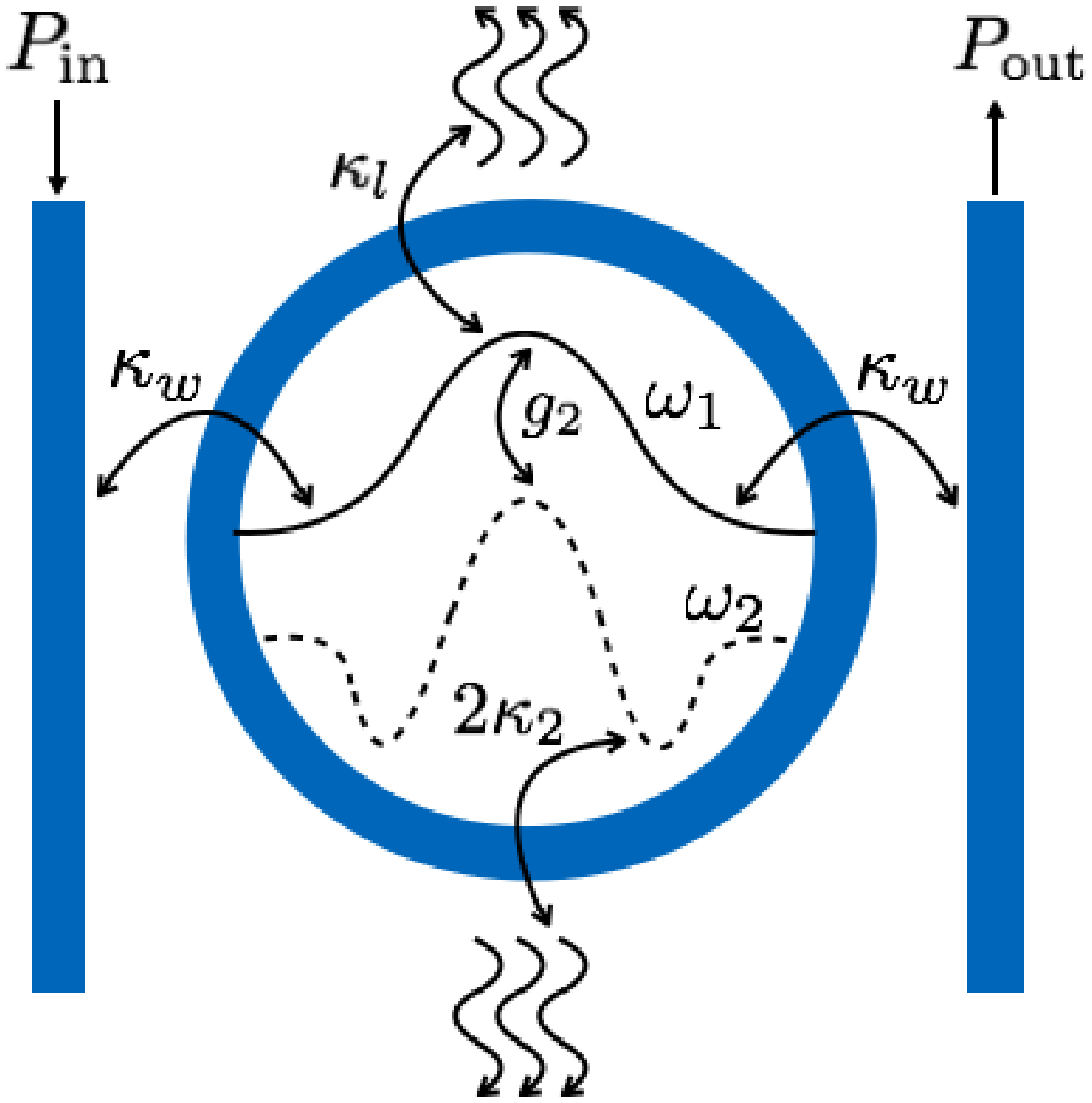}
\caption{}
\label{single_cav}
\end{subfigure}
\begin{subfigure}{0.32\textwidth}
\includegraphics[scale=0.21]{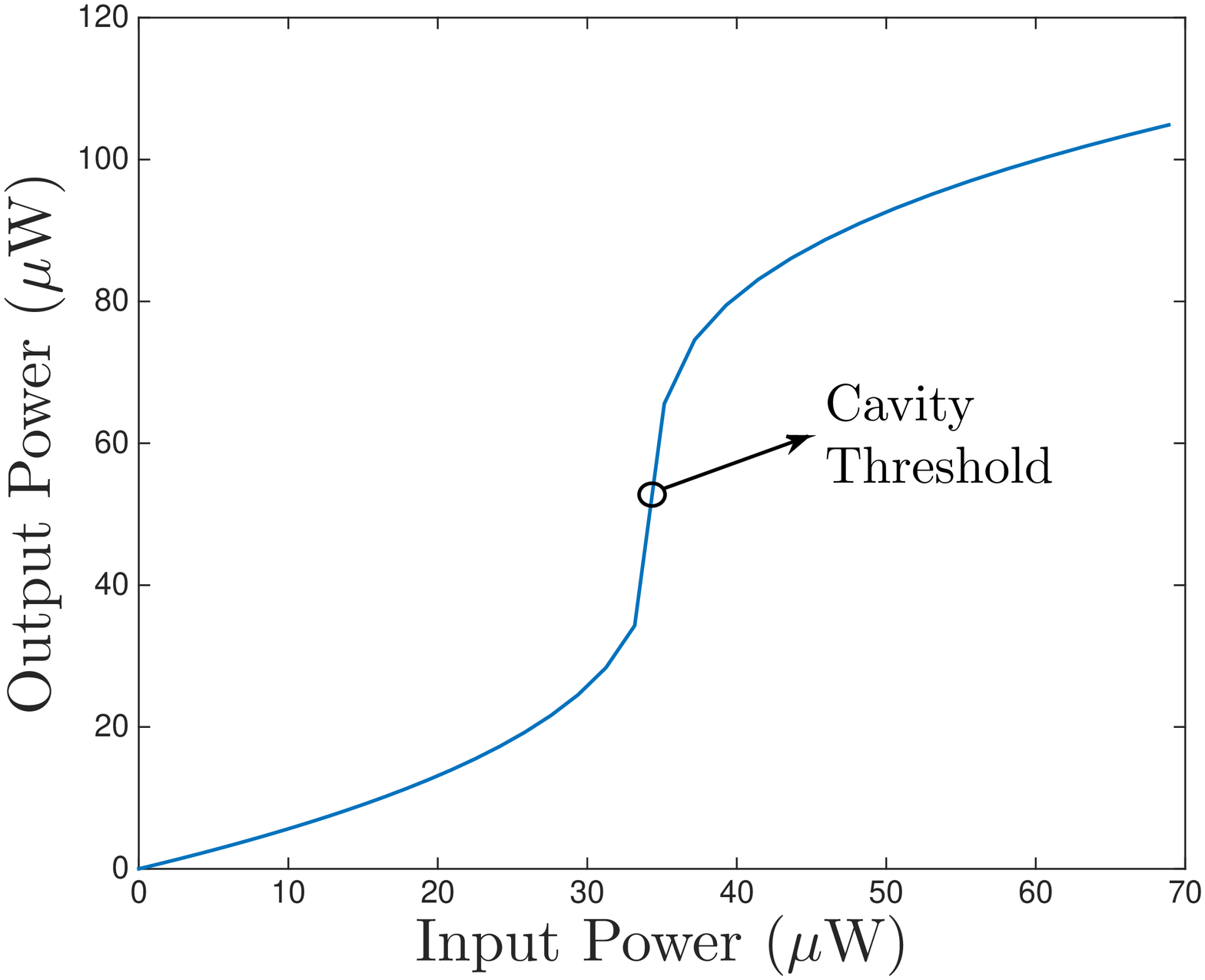}
\caption{}
\label{steady_state}
\end{subfigure}
\begin{subfigure}{0.32\textwidth}
\includegraphics[scale=0.21]{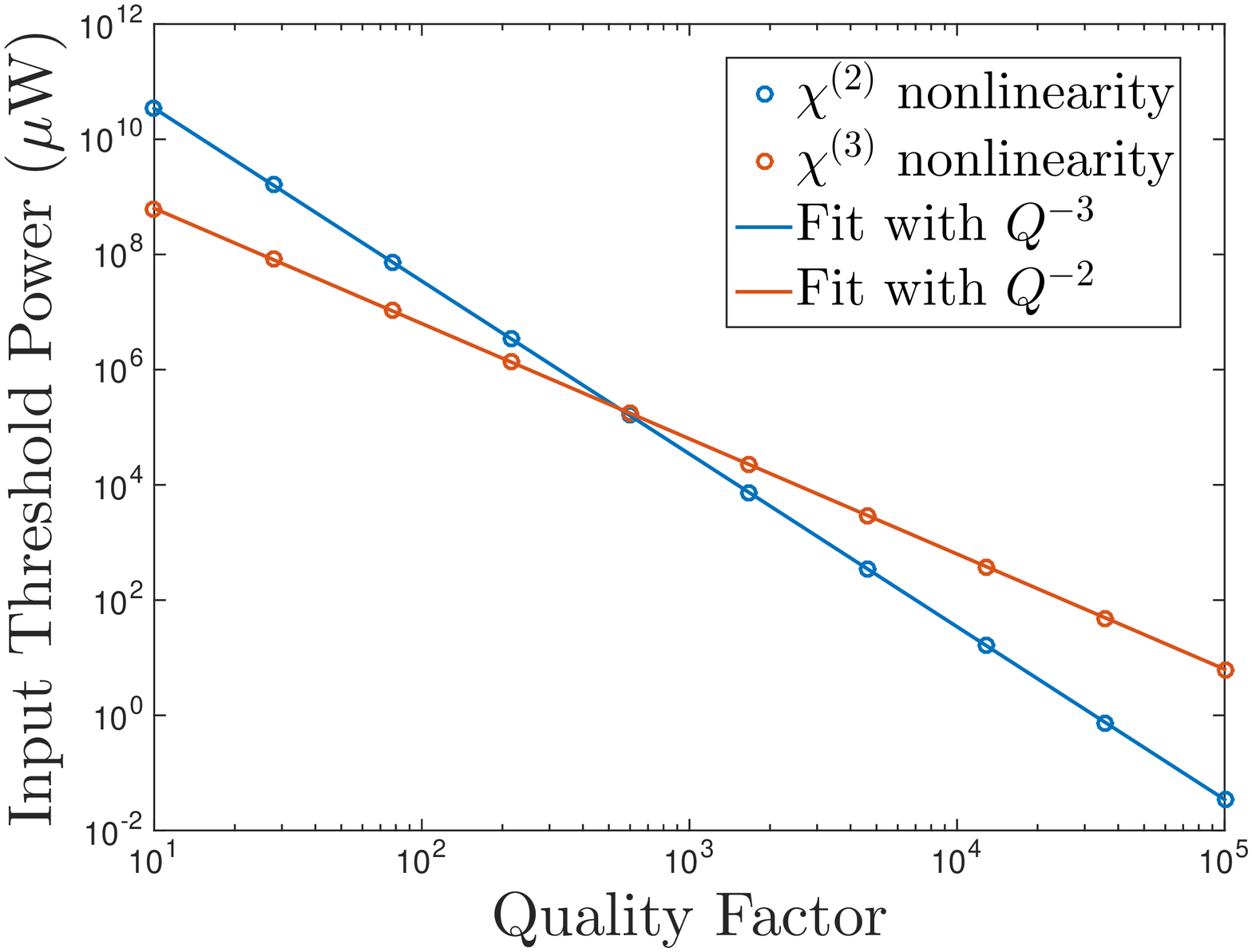}
\caption{}
\label{thresh_vs_Q}
\end{subfigure}
\caption{(a) Schematic of a single bimodal cavity coupled to two waveguides that serve as the input and output ports. The two modes (fundamental mode at $\omega_1$ and second harmonic mode at $\omega_2$) are coupled via the $\chi^{(2)}$ nonlinearity with an effective coupling constant $g_2$  (b) Steady state characteristics of a single cavity with $\chi^{(2)}$ nonlinearity for $Q \sim 10^4$ (c) Variation of the input threshold power with the cavity quality factor for both $\chi^{(2)}$ and $\chi^{(3)}$ cavities. Resonant wavelength $\lambda_0 = 2\pi c/\omega_1 = 1550$ nm, $\hbar g_2 \sim 0.5 \ \mu$eV and $\hbar g_3 \sim 30$ peV is assumed in all the simulations. }
\end{figure}

A bimodal cavity with $\chi^{(2)}$ nonlinearity couples the fundamental mode (at angular frequency $\omega_1$) and the second harmonic mode (at angular frequency $\omega_2$) through the nonlinearity. Fig. \ref{single_cav} shows the schematic of a $\chi^{(2)}$ cavity (with a nonlinear coupling constant $g_2$) coupled to an input waveguide and an output waveguide by a coupling constant $\kappa_w$. The Wigner quasi-probability distribution \cite{carter1995quantum,mandel1995optical,carmichael2009open,gardiner1985input}  can be used to model the fundamental and second harmonic cavity mode with stochastic processes $a_1(t)$ and $a_2(t)$ respectively, which satisfy the following Ito's differential equations :
\begin{subequations}
\begin{align}
\frac{\text{d}a_1(t)}{\text{d}t} &= -(i\Delta_1 + \kappa_1) a_1(t) -2ig_2 a_1^*(t) a_2(t)  - \sqrt{\kappa_w} {b}_\text{in}(t) \nonumber\\ &- \sqrt{\kappa_w} \eta_1(t)-\sqrt{\kappa_l} \ \eta_l(t) \label{eq.1} \\
\frac{\text{d}a_2(t)}{\text{d}t} &= -(i\Delta_2 + \kappa_2) a_2(t) -ig_2 a_1^2(t) - \sqrt{2\kappa_2} \eta_2(t) \label{eq.2}
\end{align}
\end{subequations}
where $\Delta_1 = \omega_1-\omega_L$ and $\Delta_2 = \omega_2 - 2\omega_L$ are the detunings of the fundamental mode frequency and the second harmonic mode frequency from the input laser frequency $\omega_L$ respectively, $2\kappa_1 = 2\kappa_w+\kappa_l$ is the net loss in the fundamental mode ($\kappa_l$ being the intrinsic loss in the fundamental mode) and $2\kappa_2$ is the net loss in the second harmonic mode. For a coherent drive, the input $b_\text{in}(t) = \bar{b}_\text{in}(t)+\eta_\text{in}(t)$, where $\bar{b}_\text{in}$ is the deterministic input signal and $\eta_\text{in}(t)$ is a gaussian white noise process. The output waveguide and loss ports of the cavity also act as sources of quantum noise \cite{santori2014quantum}, which are modeled by independant gaussian white noise processes $\eta_1(t)$, $\eta_l(t)$ and $\eta_2(t)$ respectively. All the gaussian white noise processes satisfy $\langle \eta_\text{in}^*(t+\tau)\eta_\text{in}(t)\rangle = \langle \eta_\text{1}^*(t+\tau)\eta_\text{1}(t)\rangle = \langle \eta_\text{2}^*(t+\tau)\eta_\text{2}(t)\rangle = \langle \eta_{l}^*(t+\tau)\eta_{l}(t)\rangle = \delta(\tau)/2$, $\delta(\tau)$ being the dirac-delta function \cite{gardiner2004quantum}. 

The stochastic differential equations (Eqs.~\ref{eq.1} and \ref{eq.2}) can be numerically solved using the Euler-Maruyama scheme \cite{saito1996stability,lamba2007adaptive}, wherein the gaussian white noise processes are modelled by adding a complex gaussian random variable with mean 0 and variance $(2 \delta t )^{-1}$ at every update ($\delta t$ is the sampling interval used to discretize the stochastic processes $a_1(t)$ and $a_2(t)$). Moreover, a consequence of approximating the input quantum noise by gaussian white noise processes is the presence of non-zero noise at all frequencies -- so as to obtain realistic estimates for the SNR, it is necessary to consider only the noise within the detector bandwidth. In all our calculations, we model the detector by an ideal low pass filter with $\sim 5$ GHz bandwidth.

The nonlinear coupling constant $g_2$ is a function of the field profiles of the fundamental and second harmonic mode and can be estimated via \cite{majumdar2013single}:
\begin{equation} \label{nl_cp}
\hbar g_2 \sim  \epsilon_0 \bigg(\frac{\hbar \omega_1}{\epsilon_0 n_0^2} \bigg)^{1.5} \chi^{(2)} \int e_1^2(\textbf{r})e_2^*(\textbf{r}) \text{d}^3\textbf{r}  
\end{equation}
where $e_1(\textbf{r})$ and $e_2(\textbf{r})$ are the approximate normalized scalar modal field profiles of the fundamental and second harmonic mode and $n_0$ is the cavity refractive index. Under the assumption of perfect phase matching between the two cavity modes, $\hbar g_2 \sim  \epsilon_0 ({\hbar \omega_1}/{\epsilon_0 n_0^2} )^{1.5} {\chi^{(2)}}/{\sqrt{V_2}}$, with $V_2 = (\int |e_1(\bold{r})|^3 \text{d}^3\textbf{r})^{-2}$ being the effective modal volume of the cavity. The steady state characteristics of this system i.e.~the variation of the output power $P_\text{out} =\hbar \omega_1 \kappa_w \langle|a_1|^2\rangle$ with the input power $P_\text{in} = \hbar \omega_1 |b_\text{in}|^2$ shows a nonlinear transition from a small output power to a large output power near an input threshold (Fig.~\ref{steady_state}) \cite{fryett2015cavity}. For perfectly matched detunings and quality factors i.e.~$\Delta_2 = 2\Delta_1$ and $\kappa_2 = 2\kappa_1$, the sharpest possible nonlinear transition in the output power is obtained at $\Delta_1 = (2+\sqrt{3})\kappa_1$ and the input threshold power $P_\text{th,2}$ at this detuning is given by (refer to the Appendix A for a detailed derivation):
\begin{equation}\label{thresh_Second}
 P_\text{th,2} = \frac{16(12+7\sqrt{3})}{9}  \frac{\hbar \omega_1\kappa_1^4}{g_2^2 \kappa_w} \sim \frac{V_2}{Q^3}
\end{equation} 
with $Q = \omega_1/4\pi \kappa_1$ being the quality factor of the fundamental cavity mode. Since $\kappa_1 = \kappa_w+\kappa_l/2$, it can be deduced from Eq. \ref{thresh_Second} that the input threshold is minimum when $\kappa_w = \kappa_l/6$ -- this relationship between the intrinsic loss $\kappa_l$ and the optimum cavity-waveguide coupling constant $\kappa_w$ is assumed in all the designs throughout this paper.
 A similar nonlinear input-output characteristics can also be obtained by employing a cavity with $\chi^{(3)}$ nonlinearity. The corresponding nonlinear coupling constant for the $\chi^{(3)}$ nonlinearity is given by $\hbar g_3 = 3\hbar^2 \omega_1^2 \chi^{(3)}/4\epsilon_0 n_0^4 V_3$ with $V_3= (\int |e_1(\bold{r})|^4 \text{d}^3\textbf{r})^{-1}$ being the effective modal volume \cite{ferretti2012single}. The sharpest possible transition in the steady-state characteristics is obtained for $\Delta_1 = 2g_3-\sqrt{3}\kappa_1$ and the input threshold is given by $P_\text{th,3} = 4\hbar \omega_1 \kappa_1^3/3\sqrt{3}g_3 \kappa_w \sim V_3/Q^2$, which is minimum for $\kappa_w = \kappa_l/4$. For available optically nonlinear materials ($\chi^{(2)} \sim 2\times 10^{-10} \text{m/V}$ for GaP \cite{sweet2010gaas,combrie2008gaas} and $\chi^{(3)} \sim 5 \times 10^{-19} \text{m}^2/\text{V}^2$ for Si \cite{dinu2003third,bristow2007two,niehusmann2004ultrahigh,deotare2009high}) and assuming a ring resonator cavity structure, the nonlinear coupling constants approximately evaluate to $\hbar g_2 \sim 0.5 \ \mu$eV and $\hbar g_3 \sim 30$ peV at telecom wavelengths ($\lambda_0 = 2\pi c/\omega_1 = 1550$ nm). It can also be noted that the modal volumes $V_2$ and $V_3$ appearing in Eq.~\ref{nl_cp} are different for the two nonlinearities (refer to Appendix B for their definitions and estimates for the ring-resonator structure), but this difference does not significantly effect the relative order of magnitudes of $g_2$ and $g_3$.
 
Fig.~\ref{thresh_vs_Q}. shows the variation of the input threshold $P_\text{th,2}$ and $P_\text{th,3}$ with the fundamental mode quality factor $Q$. We emphasise on the fact that while the coupling constant $g_2$ of the second order nonlinear cavity $\sim$ 4 orders of magnitude larger than the coupling constant $g_3$ of the third order nonlinear cavity, the relative magnitudes of the input thresholds also depend on the quality factor of the cavity under consideration. At very small quality factors ($Q< 600$), the third-order cavity consumes less power than the second-order cavity -- the threshold scales as $Q^{-2}$ as opposed to the $Q^{-3}$ scaling for the latter. Beyond $Q = 4(7+4\sqrt{3}) \omega_1 g_3/3\pi g_2^2 \sim 600$ (derived in Appendix A), the $\chi^{(2)}$ nonlinearity has an input threshold which becomes increasingly smaller with $Q$ as compared to the $\chi^{(3)}$ nonlinearity. For experimentally achievable quality factors ($Q \sim 10^4$), the cavity with $\chi^{(2)}$ nonlinearity is $\sim 20$ times more power efficient as compared to the cavity with $\chi^{(3)}$ nonlinearity. 

\begin{figure}[htpb]
\centering
\begin{subfigure}{0.49\textwidth}
\includegraphics[scale=0.24]{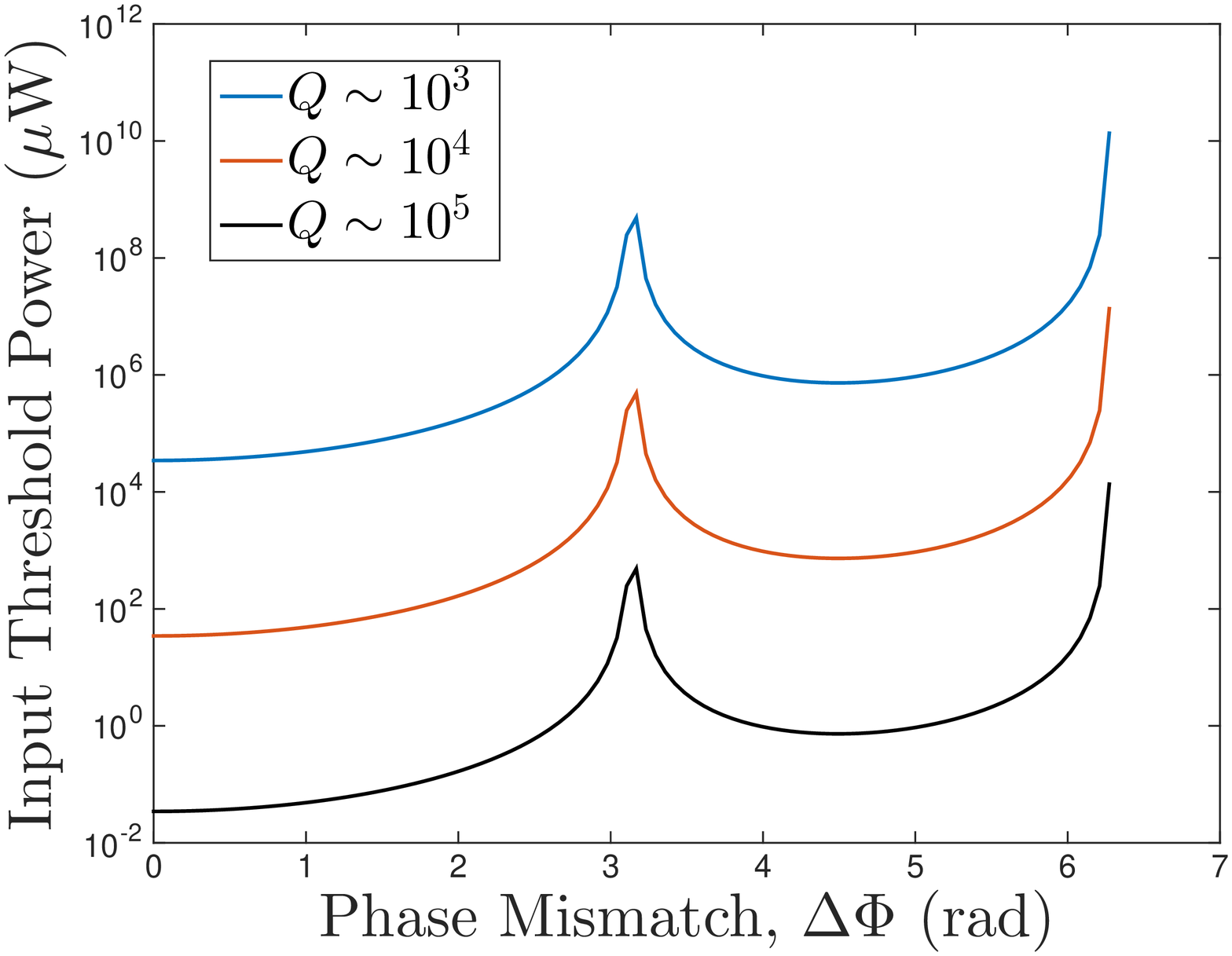}
\caption{}
\label{single_cav_pmm}
\end{subfigure}
\begin{subfigure}{0.49\textwidth}
\includegraphics[scale=0.23]{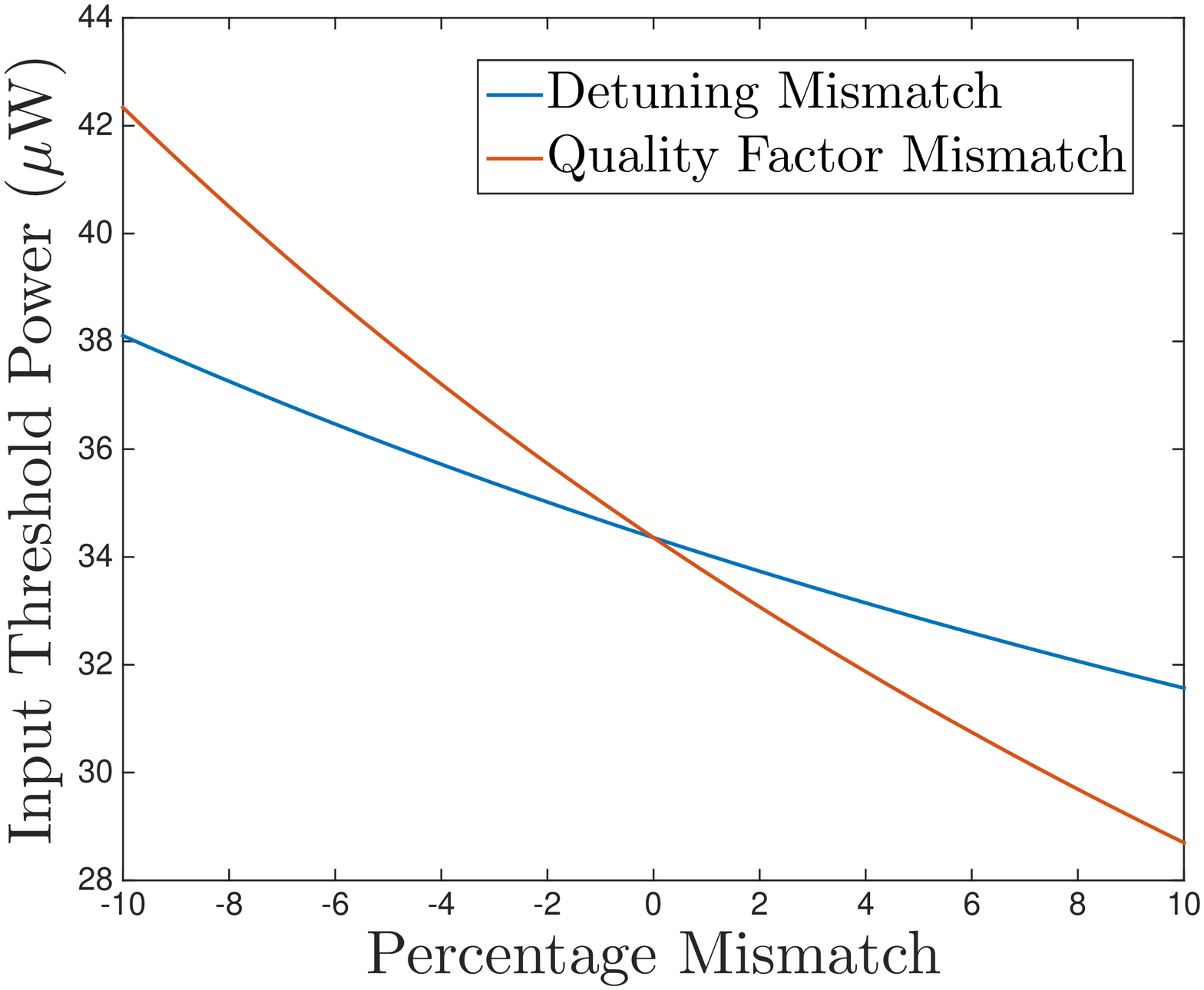}
\caption{}
\label{single_cav_mimat}
\end{subfigure}
\caption{(a) Variation of the input threshold power with the phase mismatch $\Delta \Phi$ for $\chi^{(2)}$ nonlinearity (c) Variation of the input threshold power with the detuning and quality factor mismatch for $\chi^{(2)}$ nonlinearity with $Q \sim 10^4$. Resonant wavelength $\lambda_0 = 2\pi c/\omega_1 = 1550$ nm and $\hbar g_2 \sim 0.5 \ \mu$eV (for $\Delta \Phi = 0$) is assumed in all the simulations. }
\end{figure}

Employing the $\chi^{(2)}$ nonlinearity can potentially achieve low power operation compared to the $\chi^{(3)}$ nonlinearity, however a significant disadvantage of using $\chi^{(2)}$ nonlinear cavity is the input threshold's sensitivity to mismatch between the fundamental and second harmonic mode. Phase mismatch between the two cavity modes is the most important -- to estimate its impact, we consider a ring-resonator cavity with radius $R$ and modal field profiles approximated by $e_1(\textbf{r}) \approx \alpha(r,z) \exp(i\beta_1 R \phi)$ and $e_2(\textbf{r}) \approx \alpha(r,z) \exp(i\beta_2 R \phi)$. The nonlinear coupling constant $\hbar g_2$ can then be evaluated using Eq. \ref{nl_cp} to obtain:
\begin{equation}
\hbar g_2 = \epsilon_0 \bigg(\frac{\hbar \omega_1}{\epsilon_0 n_0^2} \bigg)^{1.5} \frac{\chi^{(2)}}{\sqrt{V_2}}\bigg(\frac{ \sin \Delta \Phi}{\Delta \Phi}\bigg)
\end{equation}
where $ V_2 = [2\pi\int_{r=0}^{\infty} \int_{z=-\infty}^{\infty} \alpha^3(r,z) r\text{d}r\text{d}z]^{-2}$ is the effective modal volume and $\Delta \Phi = \pi R (2\beta_1 - \beta_2)$ is the phase mismatch. Fig.~\ref{single_cav_pmm} shows the variation of the cavity input threshold as a function of the phase mismatch and it can be seen that the input threshold for a cavity with non-zero phase mismatch is significantly larger than the input threshold for a perfectly matched cavity due to the reduced nonlinear coupling constant $g_2$. Moreover, for $\Delta \Phi$ close to $\pi$ or $2\pi$, the input threshold can increase by several orders of magnitude as compared to $\Delta \Phi = 0$. While our analysis assumes a ring-resonator geometry and a simple approximation to the modal field profiles, a qualitatively similar increase in the threshold power with phase mismatch is expected even for more complex cavity structures. 

In addition to phase mismatch, it is also difficult to experimentally achieve perfectly matched detuning and quality factor between the two cavity modes ($\Delta_2 = 2\Delta_1$ and $\kappa_2 = 2\kappa_1$). Fig.~\ref{single_cav_mimat} shows the variation of the cavity input threshold with mismatch in the detunings and quality factors (calculations outlined in Appendix A) and it can be seen that the input threshold varies significantly ($\sim 30\% $) for a $\pm 10 \%$ mismatch between the two modes. This problem of mismatch between the cavity modes does not arise in the case of $\chi^{(3)}$ nonlinearity, since it depends only on the fundamental cavity mode instead of relying on nonlinearly coupling different cavity modes. We note that, the requirement of two cavity modes with good overlap is an important and stringent trade-off while adopting the $\chi^{(2)}$ nonlinearity for reducing power consumption.

\section{Optical Logic Gates\label{sec:Optical logic networks}}

\begin{figure}[htpb]
\centering
\begin{subfigure}{0.49\textwidth}
\includegraphics[scale=0.4]{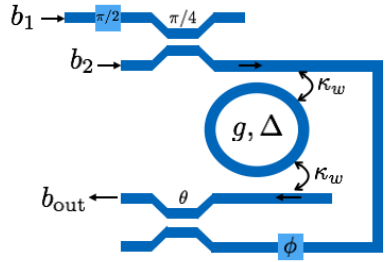}
\caption{}
\label{and}
\end{subfigure}
\begin{subfigure}{0.49\textwidth}
\centering
\includegraphics[scale=0.24]{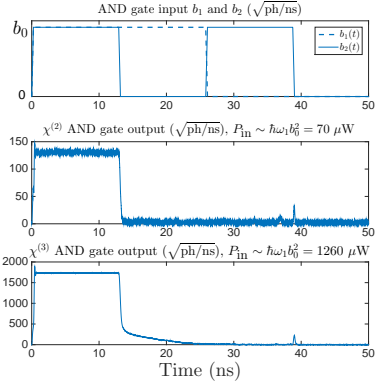}
\caption{}
\label{and_tran}
\end{subfigure}
\begin{subfigure}{0.49\textwidth}
\centering
\includegraphics[scale=0.21]{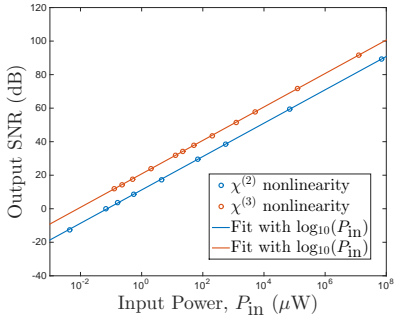}
\caption{}
\label{and_snr}
\end{subfigure}
\begin{subfigure}{0.49\textwidth}
\includegraphics[scale=0.21]{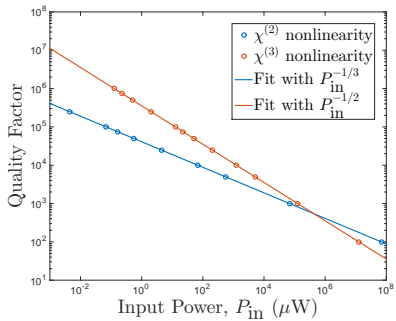}
\caption{}
\label{and_q}
\end{subfigure}
\caption{(a) Photonic circuit to implement the AND gate (b) Transient response of an AND gate implemented with both $\chi^{(2)}$ and $\chi^{(3)}$ nonlinear cavities ($Q \sim 10^4$) (c) Output SNR of the AND gate as a function of the input power (d) Cavity quality factor required to operate the AND gate as a function of the input power. For SNR calculations, it is assumed that gate output is measured with a detector of $\sim 5$ GHz bandwidth. Other Parameters: $\hbar g_2 \sim 0.5\ \mu$eV, $\hbar g_3 \sim 30$ peV and resonant wavelength $\lambda_0 = 2\pi c/\omega_1 = 1550$ nm.  }
\end{figure}

Fig.~\ref{and}. shows the photonic circuit implementing an optical AND gate \cite{santori2014quantum,mabuchi2011nonlinear}. Using a directional coupler and a phase shifter, $(b_1+b_2)/\sqrt{2}$ is fed into the cavity, where $b_1$ and $b_2$ are the inputs of the AND gate. This itself ensures that the output $b_\text{out}$ is LOW when both the inputs are LOW, thereby trivially satisfying one line in the AND gate truth table. The input high level $b_0$ (i.e. the magnitude of $b_1$ or $b_2$ when the signal represents a HIGH input) is designed so as to ensure that the cavity input is larger than the cavity threshold when both inputs are high and smaller than the cavity threshold otherwise. The parameters ($\theta,\phi$) of the directional coupler and phase shifter producing the output are chosen so as to achieve a nearly zero output for $(b_1,b_2) \equiv$ (HIGH, LOW). For the output of the AND gate to be cascadable, the output high level should be as close to the input high-level as possible. This avoids the need of cascading high-gain amplifiers with the logic gate to restore the output high level to the input high level. The input-high level is thus designed to maximize the ratio of the output high level to the input high level. 

For $Q \sim 10^4$, the transient response of an AND gate implemented with both $\chi^{(2)}$ and $\chi^{(3)}$ nonlinearities is shown in Fig.~\ref{and_tran}. Clearly, at quality factors of this order of magnitude the input power ($P_\text{in}\sim \hbar \omega_1b_0^2$) required for a $\chi^{(3)}$ based design is $\sim 20$ times larger than that required for a $\chi^{(2)}$ based design. However, we also note from Fig.~\ref{and_snr} that for designs operating at the same input power, the $\chi^{(3)}$ AND gate shows a larger SNR as compared to the $\chi^{(2)}$ AND gate. This can be attributed to the noise being added to the second harmonic mode in the $\chi^{(2)}$ nonlinear cavity ($\sqrt{2\kappa_2}\eta_2(t)$ in Eq.~\ref{eq.2}). On the other hand, the cavity quality factor required to design a $\chi^{(3)}$ AND gate operating at low input power is significantly higher as compared to those required by the $\chi^{(2)}$ AND gate. For instance, it can be seen from Fig.~\ref{and_q} that the $\chi^{(2)}$ AND gate designed with $Q \sim 10^4$ can operate at an input power of $\sim 10\ \mu$W with an SNR of $\sim$ 20 dB whereas the $\chi^{(3)}$ AND gate requires $Q \sim 10^5$ to operate at similar input powers even though it offers a significantly higher SNR ($\sim 30$ dB).  

\begin{figure}[htpb]
\centering
\begin{subfigure}{0.49\textwidth}
\centering
\includegraphics[scale=0.22]{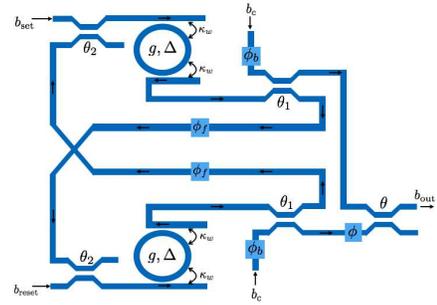}
\caption{}
\label{latch}
\end{subfigure}
\begin{subfigure}{0.49\textwidth}
\centering
\includegraphics[scale=0.23]{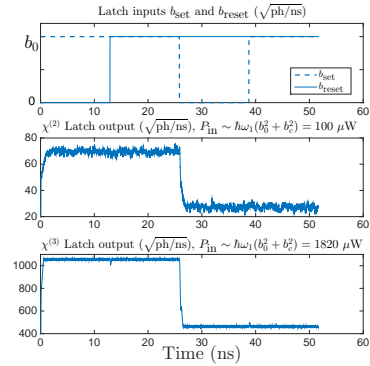}
\caption{}
\label{latch_tran}
\end{subfigure}
\begin{subfigure}{0.49\textwidth}
\centering
\includegraphics[scale=0.21]{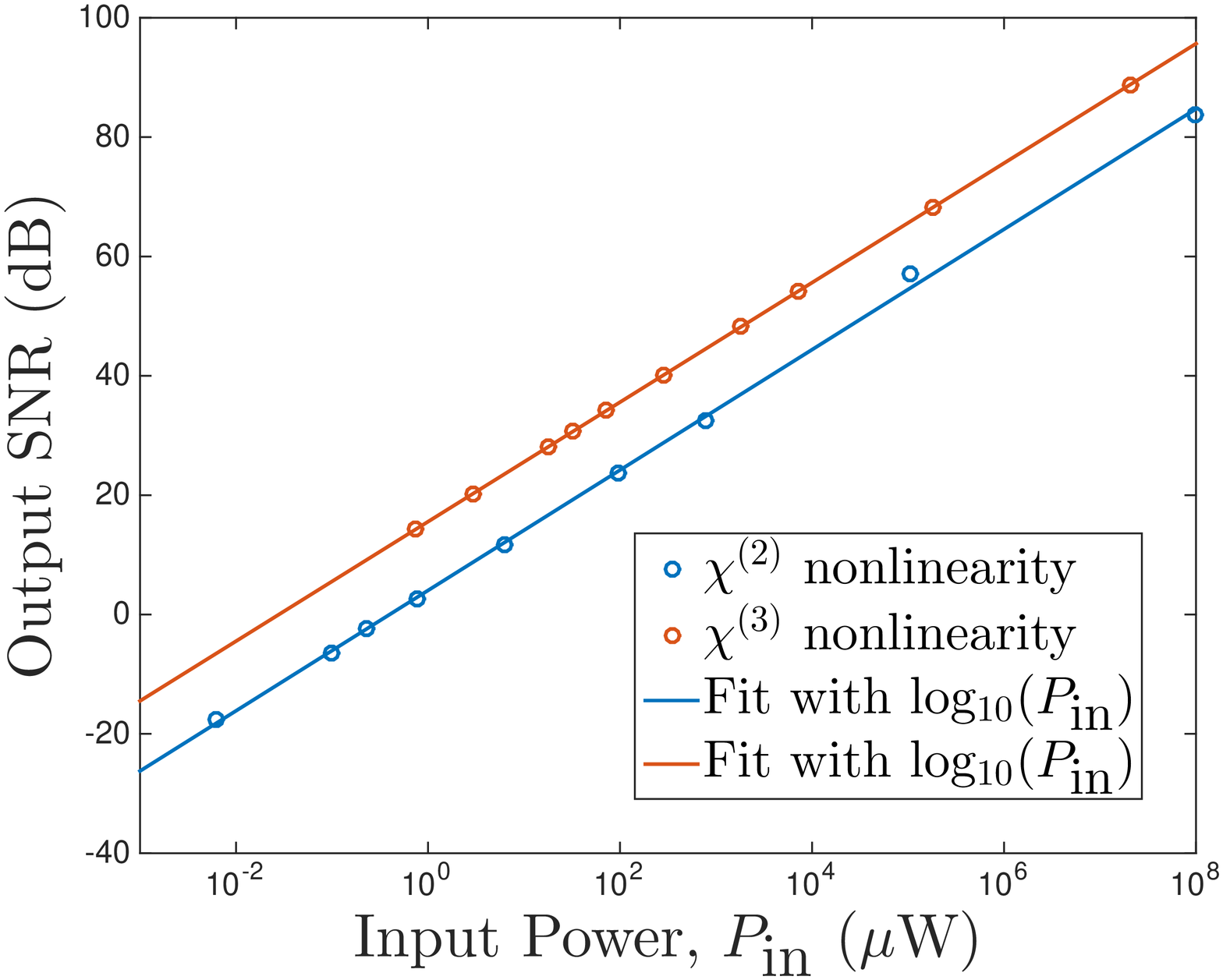}
\caption{}
\label{latch_snr}
\end{subfigure}
\begin{subfigure}{0.49\textwidth}
\centering
\includegraphics[scale=0.21]{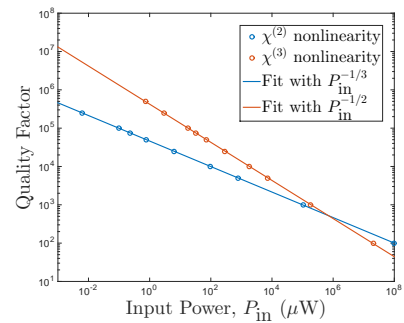}
\caption{}
\label{latch_q}
\end{subfigure}

\caption{(a) Photonic circuit for implementing the latch (b) Transient response of the latch implemented with both $\chi^{(2)}$ and $\chi^{(3)}$ nonlinearity ($Q \sim 10^4$) (c) Output SNR of the latch as a function of the input power (d) Cavity quality factor required to operate the latch gate as a function of the input power. For SNR calculations, It is assumed that the circuit output is measured with a detector of $\sim 5$ GHz bandwidth. Other parameters: $\hbar g_2 \sim 0.5\ \mu$eV, $\hbar g_3 \sim 30$ peV and resonant wavelength $\lambda_0 = 2\pi c/\omega_1 = 1550$ nm.}
\end{figure}

A bistable latch can be designed by combining two cavities in feedback (Fig.~\ref{latch}) \cite{santori2014quantum,mabuchi2011nonlinear}. The latch state is governed by two inputs $b_\text{set}$ and $b_\text{reset}$ -- if $(b_\text{set},b_\text{reset}) \equiv (\text{HIGH},\text{LOW})$, the latch output is pulled up to HIGH, if $(b_\text{set},b_\text{reset}) \equiv (\text{LOW},\text{HIGH})$ the latch output is pulled down to LOW and when $(b_\text{set},b_\text{reset}) \equiv (\text{HIGH},\text{HIGH})$ the output is held at its previous value. The parameters $\theta$ and $\phi$ of the final directional coupler and phase-shifter can be chosen so as to achieve equal output amplitudes during both a latch set and the following hold, and a latch reset and the following hold. Additionally, the circuit parameters $\theta_1,\theta_2,\phi_b,\phi_f$ and the field $b_c,b_0$ ($b_0$ being the input high level) need to be chosen so as to ensure that the overall circuit exhibits bistability during the hold condition ($(b_\text{set},b_\text{reset}) \equiv (\text{HIGH},\text{HIGH})$) -- this is desired since the circuit output during a hold is a function of the previous state of the latch and not dependent solely on the circuit inputs ($b_\text{set},b_\text{reset}$). To achieve this, we numerically maximize the ratio of the output magnitude when the latch holds a HIGH value to the output magnitude when the latch holds a LOW value as a function of $(\theta_1,\theta_2,\phi_b,\phi_f,b_c,b_0)$. The cascadability of the output signal also requires the phase of the output to be equal both when the latch is set and when it holds a HIGH data and when the latch is reset and when it holds a LOW data. This is an additional constraint that we impose while optimising the circuit parameters so as to achieve a bistable latch circuit.

Fig.~\ref{latch_tran} shows a transient simulation of the latch circuit designed with both $\chi^{(2)}$ and $\chi^{(3)}$ nonlinearities at $Q \sim 10^4$. At quality factors of this order of magnitude, the $\chi^{(2)}$ based design operates at an input power ($P_\text{in}\sim \hbar \omega_1(b_0^2+b_c^2)$) which is $\sim 20$  times smaller than that of the $\chi^{(3)}$ based design. As with the AND gate, we observe that the latch implemented with $\chi^{(3)}$ nonlinearity offers better noise performance compared to the latch implemented with $\chi^{(2)}$ nonlinearity (Fig.~\ref{latch_snr}) for the same input power with the tradeoff being the requirement of significantly larger quality factors (Fig.~\ref{latch_q}) -- the latch designed with $\chi^{(3)}$ nonlinearity requires $Q \sim 10^5 - 10^6$ to operate at input powers $\sim 10\ \mu$W whereas the corresponding design with $\chi^{(2)}$ nonlinearity requires $Q \sim 10^4$ to operate at similar input powers. 

\section{Conclusion}
We analyze and compare the power consumption and noise performance of all-optical logic gates (AND gate, latch) implemented using networks of optical cavities with $\chi^{(2)}$ and $\chi^{(3)}$ nonlinearities. Using realistic estimates for the achievable nonlinear coupling constants, it is shown that the $\chi^{(2)}$ based photonic circuits are orders of magnitude more energy efficient as compared to the $\chi^{(3)}$ based photonic circuits at achievable quality factors ($Q\sim 10^4$). This is verified by simulating an optical AND gate and an optical latch implemented using both the nonlinearities. The quantum noise in the implemented circuits is simulated using stochastic models derived from the Wigner quasi-probability distribution function. We observe that, for the same input power, optical gates implemented with $\chi^{(3)}$ nonlinearity have larger output SNR as compared to the optical gates implemented with $\chi^{(2)}$ nonlinearity. However, optical logic gates with $\chi^{(2)}$ nonlinearity can achieve low-power operation at significantly smaller quality factors as compared to optical logic gates with $\chi^{(3)}$ nonlinearity thereby making them a more suitable candidate to reduce the overall power consumption of photonic logic circuits.
\acknowledgements
\noindent Arka Majumdar is supported by the Air Force Office of Scientific Research-Young Investigator Program under grant FA9550-15-1-0150. 
\appendix
\newpage
\section{Derivation of threshold power for a nonlinear cavity}
\noindent The input power $P_\text{in} = \hbar \omega_1 |b_\text{in}|^2$ is related to the output power $P_\text{out} = \hbar \omega_1 \kappa_w \langle|a_1|^2\rangle$ for a cavity with either $\chi^{(2)}$ or $\chi^{(3)}$ nonlinearity by a cubic equation \cite{fryett2015cavity}
\begin{equation}
P_\text{in} = \gamma_1 P_\text{out}+\gamma_2 P_\text{out}^2+\gamma_3 P_\text{out}^3
\end{equation}
where, for $\chi^{(3)}$ nonlinearity,
\begin{equation}\label{chi_3}
\gamma_1 = \frac{(\Delta_1-2g_3)^2+\kappa_1^2}{\kappa_w^2}, \gamma_2 = \frac{4g_3(\Delta_1-2g_3)}{\hbar\omega_1\kappa_w^3},\gamma_3 =  \frac{4g_3^2}{\hbar^2\omega_1^2\kappa_w^4}
\end{equation}
and for $\chi^{(2)}$ nonlinearity,
\begin{align}\label{chi_2}
&\gamma_1 = \frac{\Delta_1^2+\kappa_1^2}{\kappa_w^2}, \gamma_2 = \frac{4g_2^2 (\kappa_1 \kappa_2-\Delta_1\Delta_2)}{\hbar \omega_1\kappa_w^3(\Delta_2^2+\kappa_2^2)} \nonumber\\ &\gamma_3 = \frac{4g_2^4}{\hbar^2\omega_1^2\kappa_w^4(\Delta_2^2+\kappa_2^2)}
\end{align}
where $2\kappa_1 = 2\kappa_w+\kappa_l$ is the total loss in the cavity and is a sum of the loss through the waveguides coupled to it ($2\kappa_w$) and the intrinsic cavity loss ($\kappa_l$). It has been shown that the cavity with either nonlinearity can behave like a bistable system \cite{fryett2015cavity} for sufficiently large detuning. Since the output of a bistable cavity would depend upon its previous states, it is desirable to operate the cavity in the monostable regime. Therefore:
\begin{equation}
\frac{\text{d}P_\text{in}}{\text{d}P_\text{out}} = \gamma_1+2\gamma_2 P_\text{out}+3\gamma_3 P_\text{out}^2>0 \ \forall \ P_\text{in}, P_\text{out} > 0
 \end{equation}
 which would hold only if $\gamma_2^2 < 3 \gamma_3 \gamma_1$.
Moreover, for the sharpest possible nonlinear transition in the cavity's steady state characteristics, $\text{d}P_\text{in}/\text{d}P_\text{out} = 0$ at the input threshold $P_\text{th}$ which yields:
\begin{equation}\label{thresh}
\gamma_2^2 = 3\gamma_3 \gamma_1,  P_\text{th} = -\frac{\gamma_2^3}{27 \gamma_3^2}
\end{equation}
Also note from Eq.~\ref{thresh} that for $P_\text{th}>0$, $\gamma_2<0$. Eq.~\ref{thresh} can be specialised for the $\chi^{(2)}$ and $\chi^{(3)}$ nonlinearity using Eqs.~\ref{chi_2} and \ref{chi_3}. For $\chi^{(3)}$ nonlinearity, we obtain:
\begin{equation}\label{thresh_chi3}
\Delta_1 = -\sqrt{3}\kappa_1+2g_3,  P_\text{th,3} = \frac{4 \hbar \omega_1}{3\sqrt{3}} \frac{ \kappa_1^3}{\kappa_w g_3} 
\end{equation}
and for $\chi^{(2)}$ nonlinearity,
\begin{equation}\label{thresh_chi2}
\frac{\Delta_1 \Delta_2}{\kappa_1 \kappa_2}-1 = \sqrt{3} \bigg(\frac{\Delta_1}{\kappa_1}+\frac{\Delta_2}{\kappa_2} \bigg), P_\text{th,2} = \frac{4 \hbar \omega_1}{27 g_2^2} \frac{(\Delta_1 \Delta_2 - \kappa_1 \kappa_2)^3}{\kappa_w(\Delta_2^2+\kappa_2^2)}
\end{equation}

In our analysis of the optical logic gates, we have assumed the fundamental and the second harmonic mode of the bimodal cavity to be perfectly matched in both detuning and modal line-width, i.e.~$\Delta_2 = 2\Delta_1$ and $\kappa_2 = 2\kappa_1$. Under this assumption, Eq.~\ref{thresh_chi2} can be simplified to obtain $\Delta_1 = (2+\sqrt{3}) \kappa_1$ and $P_\text{th,2} = {16(12+7\sqrt{3})}{\hbar \omega_1\kappa_1^4}/{9}{\kappa_w g_2^2}$. For a given intrinsic cavity loss $\kappa_l$, it can be seen that the $P_\text{th,2}$ is minimum for $\kappa_w = \kappa_l/6 = \kappa_1/4$ and $P_\text{th,3}$ is minimised for $\kappa_w = \kappa_l/4 = \kappa_1/3$. This along with the expressions for the input thresholds $P_\text{th,2}$ and $P_\text{th,3}$ allows the calculation of the quality factor at which the $\chi^{(2)}$ and $\chi^{(3)}$ have the same threshold power which evaluates to $Q = 4(7+4\sqrt{3}) \omega_1 g_3/3\pi g_2^2$. It can be noted that Eq.~\ref{thresh_chi2} is valid even for the case when the detuning and quality factor of the two cavity modes are not perfectly matched, and can be used to compute the input threshold for a given percentage mismatch between the two modes.

\section{Modal volume calculation}
\noindent For calculating the modal volumes $V_2$ and $V_3$ for a ring-resonator structure, we assume scalar gaussian approximations $e_1(\bold{r})$ and $e_2(\bold{r})$ to the modal fields at frequency $\omega_1$ and $\omega_2$ respectively ($p \in \{1,2\})$:
\begin{equation}\label{fld_pro}
e_p(\textbf{r}) = \frac{\exp(i\beta_p R\phi)}{\sqrt{2\pi^2\sigma_r \sigma_z R}} \exp\bigg[-\frac{1}{2}\bigg(\frac{(r-R)^2}{\sigma_r^2} +\frac{z^2}{\sigma_z^2} \bigg) \bigg]
\end{equation}
where $R$ is the mean radius of the ring resonator, $\beta_p = \omega_p n_0/c$ and $\sigma_r$, $\sigma_z$ are measures of the modal confinement in the radial and $z$ direction respectively. For simplicity, we have also assumed the intensity profile $|e_p(\textbf{r})|^2$ to be independant of the modal frequency. Since $e_p(\textbf{r})$ is a periodic function of $\phi$, the constants $\beta_p$ satisfy:
\begin{equation}\label{res}
\beta_p \approx \frac{\omega_pn_0}{c}  = \frac{m_p}{R}, \ \forall \ p\in\{1,2\}, \text{for some}\ m_p \in \mathbb{N}
\end{equation}
where $n_0$ is the refractive index of the material used in the cavity. This is equivalent to $R = m_1 c/\omega_1 n_0 = m_2 c/\omega_2 n_0$. With available fabrication facilities, the smallest possible $\sigma_r$ and $\sigma_z$ that can be achieved are of the order of $ \lambda_0/2n_0 = \pi c/n_0 \omega_1$. The physical structure of the ring-resonator constraints $R$ to be greater than $2\sigma_r\sim2\pi c/\omega_1n_0$ which, with Eq.~\ref{res}, implies $R \sim 7c/\omega_1 n_0$. For a $\chi^{(2)}$ nonlinearity, the modal volume is given by \cite{majumdar2013single}
\begin{equation}
V_2 = \bigg[\int_{r=0}^{\infty} \int_{\phi = 0}^{2\pi} \int_{z=-\infty}^{\infty} |e_1(\bold{r})|^3 r\text{d}r\text{d}\phi\text{d}z\bigg]^{-2}
\end{equation}
Using Eq.~\ref{fld_pro}, the modal volume evaluates to $V^{(2)} \approx 9\pi^2 \sigma_z \sigma_r R/2 \sim 63 \pi\lambda_0^3/16n_0^3$. For a $\chi^{(3)}$ nonlinearity, the modal volume can be calculated from \cite{ferretti2012single}
\begin{equation}
V_3 = \bigg[\int_{r=0}^{\infty} \int_{\phi = 0}^{2\pi} \int_{z=-\infty}^{\infty} |e_1(\bold{r})|^4 r\text{d}r\text{d}\phi\text{d}z\bigg]^{-1}
\end{equation}
which approximately evaluates to $V^{(3)} \approx 4\pi^2 \sigma_z \sigma_r R \sim 7 \pi \lambda_0^3/2n_0^3$. It is to be noted that these estimates for modal volumes are lower bounds on the volumes that can be achieved experimentally, and are approximately of the same order of magnitude for both the nonlinearities.
\bibliographystyle{unsrt}
\bibliography{Second_order_nonlinear_logic.bib}
\end{document}